\documentclass[10pt,letterpaper,twocolumn]{article}
\usepackage{ol2}
\usepackage{amsmath}

\begin{document}
\twocolumn[
\title{Coherent 455\,nm beam production in a cesium vapor}
\author{J.T. Schultz$^{1,*}$, S. Abend$^{1,3}$, D. D\"{o}ring$^{1}$, J.E. Debs$^{1}$, P.A. Altin$^{1}$, J.D. White$^{2}$, N.P. Robins$^{1}$, and J.D. Close$^{1}$}

\address{
$^1$ARC Centre of Excellence for Quantum-Atom Optics, \\ Department of Quantum Science, Australian National University, Canberra, ACT 0200, Australia.\\
$^2$Department of Physics, Juniata College, Huntingdon, PA 16652, USA. \\
$^3$Leibniz Universit\"{a}t Hannover, Welfengarten 1, 30167 Hannover, Germany.\\
$^*$Corresponding author: justin.t.schultz@gmail.com
}
\date{\today}
\begin{abstract}
We observe coherent, continuous wave, 455\,nm blue beam production via frequency up-conversion in cesium vapor. Two infrared lasers induce strong double-excitation in a heated cesium vapor cell, allowing the atoms to undergo a double cascade and produce a coherent, collimated, blue beam co-propagating with the two infrared pump lasers.  
\end{abstract}
\ocis{270.1670, 190.7220, 020.1670,020.4180}
]
\maketitle

\noindent Effects of atomic coherence and interference have become useful and important tools in optical physics, particularly for the enhancement of nonlinear interactions between atoms and light \cite{Interf}. Utilization of these effects has led to the recent experimental realizations of quantum interference effects such as lasing without inversion (LWI) \cite{LWI}, four-wave mixing (FWM) \cite{FWM, PFWM}, electromagnetically induced transparency (EIT) \cite{EITFirst}, both fast and slow light propagation \cite{SlowFastLight}, and high-harmonic generation \cite{HighHarGen}. Examples of EIT, FWM, and LWI have been observed in rubidium \cite{EITinRb,FWMinRb,LWIinRb} and  in cesium \cite{EITinCs, FWMinCs, LWIinCs}. There have been extensive studies of the transitions in rubidium and cesium \cite{FluLock, OptCasPump} that appear suitable for frequency up-conversion, and specifically the generation of short-wavelength laser beams. 
\begin{figure}[b]
   \centering
   \includegraphics[height=1.75in]{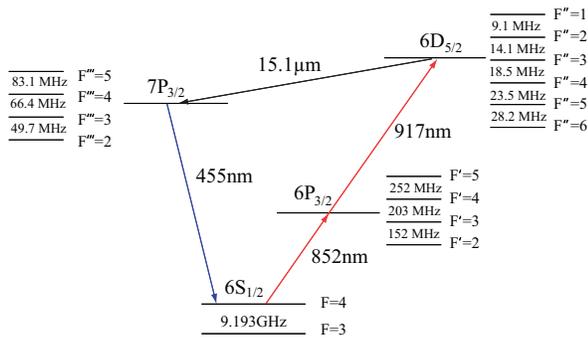}
   \caption{Energy level scheme in $^{133}$Cs. The atoms undergo a double cascade through the $7P_{3/2}$ state, producing a 455\,nm beam on the transition to the ground state.}
   \label{LevelScheme}
\end{figure}

In this Letter, we demonstrate that efficient frequency up-conversion can be achieved in resonant atomic media with low power, continuous wave lasers via techniques of interference and atomic coherence \cite{zibrov}; more specifically, we demonstrate the first realization of continuous blue beam production in cesium vapor. This work is similar to frequency up-conversion experiments in rubidium that also utilize a resonant cascade system. Zibrov {\it et al}. \cite{zibrov} and Meijer {\it et al}. \cite{unimelb} have observed and studied coherent blue beam production in rubidium vapor; they have described this process as being a result of resonant wave-mixing and lasing without inversion, respectively.  

Frequency up-conversion in a multiple-wave mixing process arises from a third order optical nonlinearity, $\chi^{(3)}$ in a medium and is strongly dependent on phase matching between the optical waves involved. That is, the phase mismatch ${\bf \delta k}= \sum_i {\bf k}_i -\sum_j {\bf k}_j$ must be zero \cite{PFWM}.  Here, ${\bf k}_i$ are the wave vectors of the incident waves and ${\bf k}_j$ are the wave vectors of the produced waves modified by the index of refraction of the nonlinear medium at the corresponding wavelength.  Phase matching can easily be achieved when the input waves have approximately the same frequency, but for incident waves of different frequencies, phase matching can only be achieved by separating the beams by an appropriate angle. 

Lasing without inversion is a process in which the quantum mechanical probability amplitudes of two (or more) pathways from the ground state to an excited state interfere destructively and suppress absorption of resonant light.  However, if an atom is in the excited state, the probability amplitudes of the two (or more) pathways from the excited state to the ground state interfere constructively, allowing for gain on that transition without a large excited state population \cite{LWI}.

 The present demonstration uses a double cascade in cesium which can be seen in Figure \ref{LevelScheme}. Two pump lasers create a strong coherence \cite{note1} on the dipole-forbidden $6S_{1/2}\rightarrow 6D_{5/2}$ transition. A small number of atoms decay via the $7P_{3/2}$ state and emit a beam of 455\,nm radiation. Through spontaneous emission, only $0.4\%$ of the atoms undergo this final cascade through the $7P_{3/2}$ state.  The steady-state solution to a four-level Optical Bloch Equation (OBE) model shows that there is no population inversion on the $7P_{3/2}\rightarrow 6S_{1/2}$ transition.  
\begin{figure}[t]
   \centering
   \includegraphics{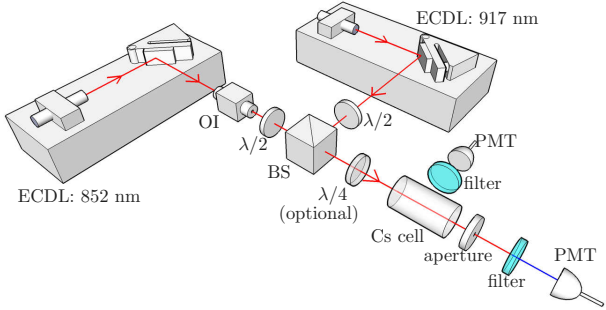}
   \caption{Schematic of experimental arrangement.  Two lasers (852\,nm, 917\,nm) excite $^{133}$Cs on the 6S-6P-6D transitions, and coherent 455\,nm radiation is emitted. The blue beam is sent through an aperture and a bandpass filter and is detected with a photomultiplier tube (PMT). A PMT is also used to measure the blue fluorescence in the cell. OI is an optical isolator, and BS is a non-polarizing beam splitter.}
   \label{SetUp}
\end{figure} 

 Figure \ref{SetUp} shows a schematic of the experimental setup. Two pump laser beams of relatively low power (30\,mW) co-propagate through a heated cesium vapor cell ($80-110\,^{\textrm{o}}$C, length 7\,cm). The data in this paper was collected when the pump beams are focused through the vapor cell to a waist of 0.3\,mm with a $f=1$m lens (corresponding to an intensity of 85\,W/cm$^{2}$); however, the focusing lens is not essential for this process.  The beams are generated with external-cavity diode lasers (ECDLs) and are resonant with the $6S_{1/2}\rightarrow6P_{3/2}$ (852\,nm) and $6P_{3/2}\rightarrow6D_{5/2}$ (917\,nm) transitions. The detuning for the 852\,nm pump beam is determined via saturated absorption in a separate vapor cell (not shown), relative to the $6S_{1/2} (F=4) \rightarrow 6P_{3/2} (F'=5)$ resonance.  For instances where the 917\,nm pump beam is scanned, the frequency of the 917\,nm pump beam is monitored via a Mach-Zehnder interferometer with a path length difference of $\Delta L = 1.17$\,m. The path length difference is calibrated through the cesium saturated absorption spectrum with the 852\,nm pump beam.

The atoms undergo a double cascade, presumably creating a beam on the $6D_{5/2}\rightarrow7P_{3/2}$ (15$\,\mu$m) transition which is thought to be amplified via population inversion on that transition. A blue (455\,nm) beam is created on the final $7P_{3/2}\rightarrow 6S_{1/2}$ transition.  The output is filtered to eliminate the infrared light from the pump beams and then measured with a photomultiplier tube. Simultaneously, the blue fluorescence of the cell is measured.

When using a vapor cell, a change in the temperature of the cell corresponds to a change in the vapor pressure and therefore, a change in the optical depth (OD) of the medium. The cell temperature and OD at 852\,nm are related via:
$$OD = N\sigma_{r} l = {p(T)\sigma_{r} l \over kT},$$
where N is the atomic density calculated from the ideal gas law, $\sigma_{r}$ the resonant cross section for circularly polarized 852\,nm light calculated from \cite{LWIinRb} and l = 7\,cm the length of the cell. The pressure $p(T)$ can be calculated using the results of Taylor and Langmuir \cite{csVaporPressure}.   The beam is observed to have the greatest power at temperatures when the majority of the 852\,nm pump beam is absorbed in the cell. 
\begin{figure}[b]
   \centering
   \includegraphics[height=2in]{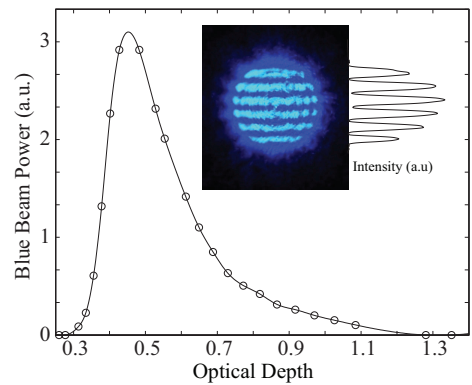}
   \caption{Measurement of blue beam intensity as a function of optical depth at 852\,nm. Interference fringes from a Mach-Zehnder interferometer with unequal path lengths (inset). Even at relatively low power ($\approx 100$\,nW) the blue pixels in the image are saturated.  A fringe visibility of $\approx 93\% $ was calculated by integrating over intensity of the green pixels in rows of the RGB image of the fringe pattern. The path length difference in the interferometer is $\Delta L = 25$\,cm.}
   \label{OD}
\end{figure}
As shown in Figure \ref{OD} we do not observe a blue beam for an OD below 0.3.  The greatest power can be achieved near an OD of 0.5, which corresponds to a temperature of around $369$\,K in our case.  For higher temperatures the blue output decreases until the atomic medium is too optically dense for the blue beam to transmit.  As the temperature of the cell increases above the peak temperature of $369$\,K, the peak blue output occurs when the 852\,nm pump beam is red-detuned from resonance. This behavior describes the observed asymmetry in Figure \ref{OD}.

Just as Zibrov {\it et al}. \cite{zibrov}, we note that a blue output power of $P_{455}\approx 4\,\mu$W corresponds to a conversion efficiency which exceeds that of typical nonlinear crystals by many orders of magnitude for the continuous wave, low power regime, and the power output could potentially be increased with the use of a build-up cavity \cite{BuildupCavity}.  

The 455\,nm beam is found to be collimated with a divergence less than 0.1\,mrad. An interference pattern from a Mach-Zehnder interferometer (Figure \ref{OD}, inset) with unequal path lengths ($\Delta L =25$\,cm) confirms that the beam has substantial spatial and temporal coherence with a fringe visibility of 93\%. The greatest efficiency is obtained when both pump beams are circularly polarized.  Kargapol'tsev {\it et al}. \cite{OptCasPump} have suggested that input beams of the same circular polarization would be the most efficient method for making a laser on the 455\,nm transition. Due to cycling transitions at each of the two excitation stages, combined with large relative transition strengths, the atoms involved in the blue beam production will predominately undergo the $6S_{1/2} (F=4)\rightarrow 6P_{3/2} (F'=5)\rightarrow 6D_{5/2} (F''=6)\rightarrow7P_{3/2} (F'''=5)\rightarrow 6S_{1/2} (F=4)$ transition, which eliminates optical pumping to the $F=3$ ground state. The $F=3$ state can still be repopulated through atom-atom collisions and collisions with the walls of the vapor cell.   For linearly polarized pump beams (or pump beams with opposite circular polarization) which couple the $F=4$ hyperfine ground state, the atoms can be excited to many different magnetic substates, and a larger number can emit to the dark $F=3$ hyperfine ground state. 


\begin{figure}[t]
   \centering
   \includegraphics[height=1.40in]{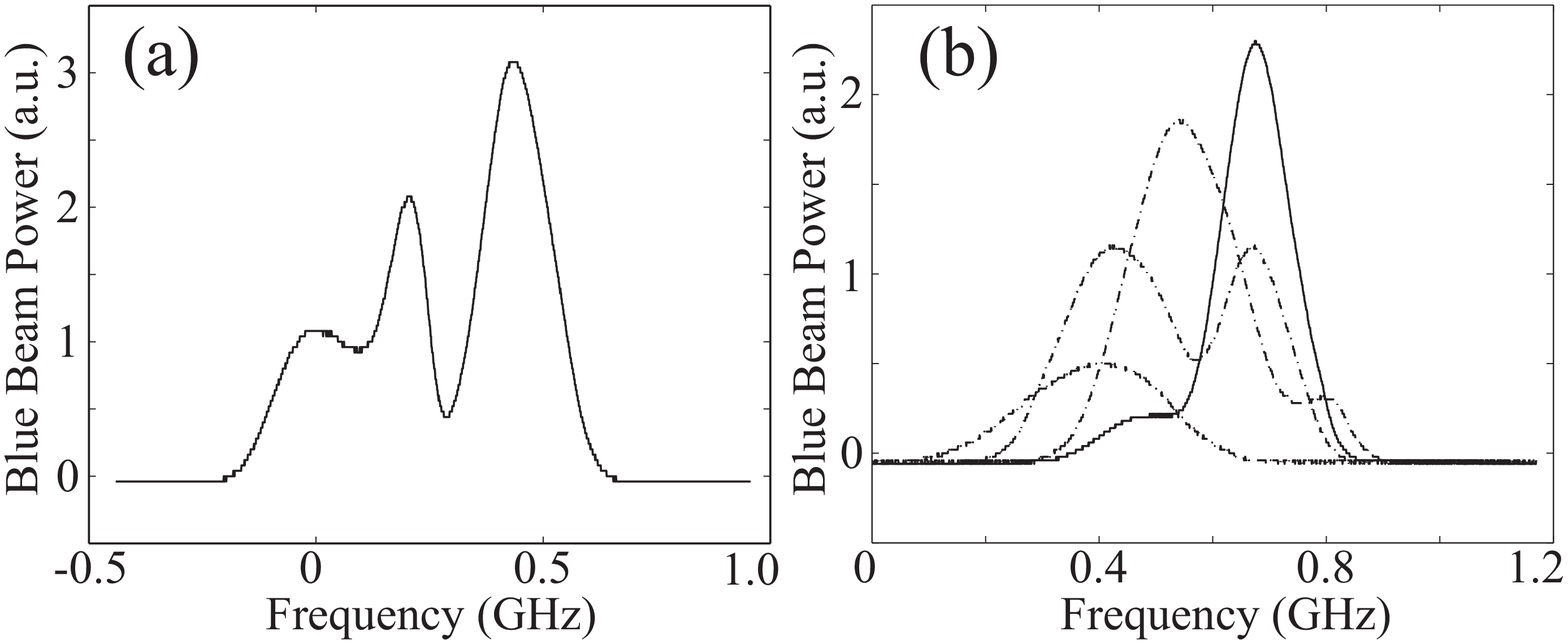}
   \caption{(a) Blue beam output as the frequency of the 917\,nm beam is scanned. The spacing between the peaks is 203\,MHz and 252\,MHz, which is the spacing of the hyperfine levels of the $6P_{3/2} $ state ($F'=3\rightarrow F'=4$ and $F'=4\rightarrow F'=5$, respectively). (b) Blue beam output for different detunings of the 852\,nm pump beam as the frequency of the 917\,nm beam is scanned. The blue output was seen for a total detuning range of about 750\,MHz for the 917\,nm pump beam, provided the 852\,nm beam was tuned to match the two-photon resonance condition.  The solid line show the peak blue output and dotted lines represent output profiles for different 852\,nm pump detunings. The frequency axis is calibrated by sending a portion of the 917\,nm beam through an interferometer with a path length difference $\Delta L = 1.17$\,m, giving a fringe separation of $\approx256$\,MHz. }
   \label{beam917}
\end{figure}

   The blue beam production was only observed when the 852\,nm pump beam coupled the $F=4$ hyperfine ground state, but was seen for a wide range ($\approx 750$\,MHz) of detunings for both input beams.  Figure \ref{beam917} shows the range of blue output observed at fixed frequencies of the 852\,nm pump beam while scanning the frequency of the 917\,nm beam.  When the 852\,nm laser coupled the $F=3$ ground state, intense fluorescence was produced, but no beam was observed. For both circularly and linearly polarized light which couples the $F=3$ ground state, a portion of the $6S_{1/2}(F=3)$ population is transferred to the dark $6S_{1/2}(F=4)$ state. Because of a relatively high transition probability to the $F=4$ state, the beam resulting from emission into the $F=3$ state would be relatively weak and presumably more easily absorbed in the vapor cell.

In conclusion, we have demonstrated frequency up-conversion in cesium on the $7P_{3/2}\rightarrow 6S_{1/2}$ transition when the two pump lasers are tuned near the $6S_{1/2}(F=4)\rightarrow6P_{3/2}\rightarrow6D_{5/2}$ transition. The beam production is most efficient when the infrared pump beams are circularly polarized, thereby utilizing cycling transitions at each excitation stage. Unlike a similar system in rubidium, no blue beam production is observed for large detunings of the pump beams. The addition of another laser to pump atoms from the $F=3$ hyperfine ground state  and a build up cavity may help increase 455\,nm beam production regardless of the polarization of the pump beams.  Although work has been done to model these systems with the Optical Bloch Equations  \cite{unimelb}, further work needs to be done to distinguish the true nature of the process responsible for the beam production, especially since the  OBE model makes no predictions about the directionality of the blue radiation.  It is possible that this method for frequency up-conversion can be extended to other elements such as sodium and lithium in order to produce a coherent beam of ultraviolet radiation (330\,nm and 323\,nm, respectively). 
The creation of the 455\,nm beam could be a promising way to study the hyperfine structure of the $7P_{3/2}$ manifold of cesium if the frequency of the beam were modulated with an acousto-optic modulator. 

This work has been supported by the Australian Research Council and  National Science Foundation (award No. PHY-0653518). JTS is thankful for a Fulbright Postgraduate Fellowship. SA thanks the Studienstiftung des Deutschen Volkes for his scholarship.


\begin{thebibliography}{30}
\bibitem{note1} The steady-state solution to a four-level OBE model shows that the coherence between the $6S_{1/2}$ and $6D_{5/2}$ states is nonzero. The off-diagonal density matrix element $\rho_{13}$ is proportional to the coherence and does not decay to zero in the steady state.
\end{thebibliography}
\end{document}